\documentclass{PoS}

\renewcommand{\logo}{\raisebox{-5mm}{\hbox{}}}

\makeatletter
\global\setbox\PoSpaper@url\hbox{}
\makeatother

\title{Search for neutrinoless double-beta decays in Ge-76 in the LEGEND
experiment}

\ShortTitle{Search for neutrinoless double-beta decays in Ge-76 in the LEGEND
experiment}

\author{\speaker{Jordan Myslik} for the LEGEND Collaboration\\
        Nuclear Science Division, Lawrence Berkeley National Laboratory, Berkeley, CA, USA \\
        E-mail: \email{jwmyslik@lbl.gov}}

\abstract{The search for neutrinoless double-beta decay is the most sensitive
technique to establish the Majorana nature of neutrinos. Two operating
experiments that look for such decays in $^{76}$Ge -- GERDA and the
\textsc{Majorana Demonstrator}
-- have achieved the lowest backgrounds and the best energy resolution in the
signal region. These are two of the most important detector characteristics for
sensitive searches of this undiscovered decay. The Large Enriched Germanium
Experiment for Neutrinoless Double Beta Decay (LEGEND) Collaboration has
formed to pursue a tonne-scale $^{76}$Ge experiment that integrates the best
technologies from these two experiments and others in the field. The
Collaboration is developing a phased experimental program that uses existing
resources as appropriate to expedite physics results, with the ultimate
discovery potential at a decay half-life beyond $10^{28}$ years. In these
proceedings, we will present the physics case, R\&D efforts and 
implementation strategies of the LEGEND experiment.}

\FullConference{The 39th International Conference on High Energy Physics (ICHEP2018)\\
		4-11 July, 2018\\
		Seoul, Korea}

\begin{document}

\section{Introduction}
Neutrinoless double-beta decay violates lepton number conservation by two units, and is a
key feature of multiple neutrino mass generation models. It would be observed
as a peak at the Q-value of an isotope's double-beta decay spectrum. For
$^{76}$Ge, sensitivity to a half-life $>10^{28}$~y in the light neutrino exchange model
\cite{Rodejohann:2011mu} would probe the entire neutrino mass 
inverted ordering (IO) parameter space and most of the normal ordering (NO) 
parameter space
\cite{Agostini:2017jim}, making it a reasonable target for a future experiment. With a signal of only 
$\sim 0.5$~counts/t$\cdot$y at a half-life of
$10^{28}$~y, ultra-low backgrounds, a large fiducial mass, and a long counting
time are required.

The \textsc{Majorana Demonstrator} \cite{Abgrall:2013rze} and GERDA (the GERmanium
Detector Array) \cite{Ackermann:2012xja} are the two currently operating experiments using
germanium crystal detectors enriched to $\sim88$\% in $^{76}$Ge (29.7~kg for
the \textsc{Majorana Demonstrator} and 37.6~kg for GERDA).  Both are located
underground to reduce cosmic ray muon flux, at the Sanford Underground Research
Facility (SURF) and the Laboratori Nazionali del Gran Sasso (LNGS),
respectively.  The \textsc{Majorana Demonstrator}'s approach of two vacuum cryostats
within a passive shield, with emphasis on ultra-clean components, is different
from GERDA's approach of submerging their detectors in a liquid argon active
shield.  These approaches both resulted in lower backgrounds and better
energy resolution than all other neutrinoless double-beta decay experiment
technologies (GERDA holding the background record of
$\sim3$~counts/(FWHM$\cdot$t$\cdot$y), and the \textsc{Majorana Demonstrator}
holding the energy resolution record of 2.5~keV FWHM at the 2039-keV Q-value).

\section{LEGEND}
Using lessons learned from GERDA, the \textsc{Majorana Demonstrator}, and
contributions from other groups, the LEGEND (Large Enriched Germanium
Experiment for Neutrinoless $\beta\beta$ Decay) collaboration plans to develop
a $^{76}$Ge-based double-beta decay experimental program, proceeding in phases
to a discovery potential at a half-life beyond $10^{28}$ years.  Existing
resources will be used as appropriate to expedite physics results.

The 200-kg first phase, LEGEND-200, will modify existing GERDA infrastructure
at LNGS, permitting early science (physics data
expected in 2021) with a world-leading experiment (sensitivity greater than
$10^{27}$~years with 1~t$\cdot$y of exposure). 
Between GERDA, the \textsc{Majorana \mbox{Demonstrator}}, 
and dedicated test stands, reducing backgrounds to the goal of less
than 0.6~counts/(FWHM$\cdot$t$\cdot$y) has already been demonstrated as
feasible.

New infrastructure will house the subsequent stages up to the 1000~kg
LEGEND-1000, over a timeline connected to the U.S. Department of Energy
down-select process for the next generation neutrinoless double-beta decay
experiment.  To achieve a discovery potential at a half-life greater than
$10^{28}$ years with approximately 10~t$\cdot$y of exposure, a background
rate less than 0.1~counts/(FWHM$\cdot$t$\cdot$y) is required.  The depth
required for sufficiently low cosmogenic backgrounds (e.g. $^{77\mathrm{m}}$Ge)
will impact the site choice, and is under investigation (e.g. \cite{Wiesinger:2018qxt}).

\section{Background reduction techniques}
A selection of the techniques to be used to achieve the LEGEND-200 and
LEGEND-1000 background reduction goals are discussed in this section.

\subsection{Electroformed copper}
For the most background-sensitive structural components (those closest to the
detectors e.g. mounts, inner shield) the \textsc{Majorana Demonstrator} uses
copper electroformed underground at Pacific Northwest National Laboratory and
SURF, which achieved average U and Th background rates of $\le 0.1$~$\mu$Bq/kg
each \cite{Abgrall:2016cct}.  Production underground reduces backgrounds from
cosmogenic $^{60}$Co.  Electroformed copper should improve on GERDA radiopurity
for LEGEND-200, and 37~kg currently in production at SURF is expected to be complete 
by the fall of 2019.

\subsection{Liquid argon veto}
External backgrounds are tagged using scintillation light they produce in the
GERDA liquid argon (LAr) volume \cite{Agostini:2017hit}.  
It is read out by photomultiplier tubes above and below the
array, and a wavelength-shifting-fiber shroud surrounding the array, read out
by silicon photomultipliers. An additional shroud around the central column was
recently added to increase light collection efficiency.  LEGEND-200 will use a
similar design, though the optimal fibre geometry for light collection
efficiency is under study.  Improved LAr purity and Xe doping are being studied
to improve light yield and attenuation.  More radiopure fibres and
signal amplification or digitization in LAr are also being studied.

The $^{42}$Ar present in natural Ar is a background to consider, since it decays
to a $^{42}$K ion which drifts to detectors, and beta decays to $^{42}$Ca with
Q~=~3.5~MeV.  Nylon shrouds around the columns limit $^{42}$K drift in GERDA and the
future LEGEND-200, and this background is cut with $~99$\% efficiency by pulse
shape analysis.  For LEGEND-1000 this background could be removed completely by
separating the detectors into four volumes containing underground Ar (free of $^{42}$Ar)
separated by copper walls from a larger natural Ar volume.  This design would
require 21 tons (15~m$^{3}$) of underground argon, a similar need to Darkside-20k 
\cite{Aalseth:2017fik}.

\subsection{Front-end electronics}
The resistive feedback charge sensitive preamplifier for each \textsc{Majorana
Demonstrator} detector consists of the Low-Mass Front-End (LMFE) \cite{Abgrall:2015hna} close to each detector's p+ contact (to minimize noise), and 
the warm preamplifier outside the shield. An MX-11 JFET, the feedback resistor,
and the feedback capacitor are on the LMFE, so the feedback loop runs 2.15~m out
to the warm preamplifier and back. Backgrounds due to the LMFE are kept low (it
is the most radiopure front-end in existence) through its low-mass design and
selection of ultra-clean materials (e.g. the 400~nm~sputtered amorphous Ge
feedback resistor).  The LMFE is in the baseline design for LEGEND-200, with research
and development into its performance in LAr and with longer cables ongoing.  Research
and development into an ASIC preamplifier that could be placed near
each LEGEND-1000 detector is also underway.

\subsection{Larger detectors}
Using larger detectors than the \textsc{Majorana Demonstrator} (P-type Point
Contact detectors -- PPCs -- average 0.85~kg) and
GERDA (Broad Energy Germanium detectors -- BEGes -- average 0.66~kg) would reduce the surface to volume ratio, and
require fewer cables and preamplifier front-ends per unit mass, reducing backgrounds. A new detector geometry, Inverted-Coaxial 
Point Contact (ICPC) detectors \cite{2011NIMPA.665...25C}, have similar performance
to PPCs and BEGes (excellent energy resolution and pulse-shape sensitivity) 
\cite{Domula:2017mei}, but can be much larger.  The current baseline for enriched 
detectors is 1.5-2.0~kg, with detectors around 3~kg and up to 6~kg being 
investigated. GERDA deployed 5 enriched ICPC detectors (each $\sim1.9$~kg) 
in a recent upgrade.

\section{Conclusions}
The \textsc{Majorana Demonstrator} and GERDA have demonstrated the lowest
backgrounds and best energy resolution of all neutrinoless double-beta decay
searches, but the next generation of experiments requires additional mass and
further reduced backgrounds.  Taking the best of these experiments, along with
additional R\&D, LEGEND will proceed in a phased fashion towards a tonne-scale
experiment with neutrinoless double-beta decay discovery potential at a half-life
beyond $10^{28}$ years.  The 200-kg first phase, LEGEND-200, has a background
reduction goal that is already demonstrated to be feasible, and will begin
taking data in 2021.

\bibliographystyle{JHEP}
\bibliography{ICHEP2018_Proceedings_LEGEND_Myslik}

%\begin{thebibliography}{99}
%\bibitem{...}
%....
%
%\end{thebibliography}

\end{document}